# THE PRIMORDIAL HELIUM ABUNDANCE

Manuel Peimbert

Instituto de Astronomía, Universidad Nacional Autónoma de México
México, 04510 D. F., Mexico; peimbert@astroscu.unam.mx

## ABSTRACT

I present a brief review on the determination of the primordial helium abundance by unit mass, $Y_P$. I discuss the importance of the primordial helium abundance in: (a) cosmology, (b) testing the standard big bang nucleosynthesis, (c) studying the physical conditions in H II regions, (d) providing the initial conditions for stellar evolution models, and (e) testing the galactic chemical evolution models.

## 1. INTRODUCTION

The homogeneous model of the expansion of the universe based on the general theory of relativity, now known as the Big Bang Theory (BBT), predicts that during the first four minutes, counted from the beginning of the expansion of the universe, there were nuclear reactions based on hydrogen that produced helium, and traces of deuterium and lithium. During the expansion, the temperature of the universe was decreasing and after these four minutes the temperature was no longer high enough to produce from nuclear reactions the other elements of the periodic table. Many millions of years later the first stars were formed with helium and hydrogen only, this helium is called primordial helium. The other elements of the periodic table, called heavy elements, were formed by nuclear reactions in the stellar interiors and a fraction of them was expelled later to the interstellar medium.

The formation of the elements is a key problem for understanding the evolution of the universe. In particular the formation of helium four ($^4$He) has been paramount for the study of cosmology and the chemical evolution of galaxies.

Over the years the increase in the accuracy of the helium abundance by mass, $Y$, of different objects and in the accuracy of the predictions of the primordial helium abundance by unit mass, $Y_P$ (also known as the primordial helium mass fraction), by Big Bang Nucleosynthesis (BBN) has led to a better understanding of the universe.

$Y_P$ is determined by means of an extrapolation to $Z = 0$ of the $Y$ values determined observationally of a sample of objects. Here, the usual normalization $X + Y + Z = 1$ is used, where $X$, $Y$, and $Z$ are the abundances per unit mass of hydrogen, helium, and the rest of the elements respectively. The extrapolation is traditionally done in the $Y$, $Z$ space by assuming a $\Delta Y/\Delta Z$ slope. More recently it has become common to use $\Delta Y/\Delta O$, where $O$ is the oxygen abundance per unit mass, because the $O$ value is easier to determine than the $Z$ value. Oxygen is the most abundant heavy element in the universe, the $O$ abundance in $O$-poor H II regions corresponds to about 53% of the total $Z$ value.

The determination of $Y_P$ is important for at least the following reasons: (a) it is one of the pillars of Big Bang cosmology and an accurate determination of $Y_P$ permits to test the Standard Big Bang Nucleosynthesis (SBBN), (b) the models of stellar evolution require an accurate initial $Y$ value; this is given by $Y_P$ plus the additional $Y$ produced by galactic chemical evolution, which can be estimated based on the observationally determined $\Delta Y/\Delta O$ ratio, (c) the combination of $Y_P$ and $\Delta Y/\Delta O$ is needed to test models of galactic chemical evolution, (d) to test solar models it is necessary to know the initial solar abundances, which are different to the photospheric ones due to diffusive settling, this effect reduces the helium and heavy element abundances in the solar photosphere relative to that of hydrogen, the initial solar abundances can be provided by models of galactic chemical evolution, (e) the determination of the $Y$ value in metal poor H II regions requires a deep knowledge of their physical conditions, in particular the $Y$ determination depends to a significant degree on their density and temperature distribution, therefore accurate $Y$ determinations combined with the assumption of SBBN provide a constraint on the density and temperature structure of H II regions.

According to the SBBN, the primordial abundances of the light isotopes D, $^7$Li, $^3$He, and $^4$He depend on one parameter only, the baryon-to-photon ratio. Therefore, the four abundance values provide altogether a strong constraint on the cosmological models. Unfortunately, their determination is not an easy task, and each of the four isotopes poses a different challenge. In particular, although $^4$He is the most abundant of the four and the easiest to measure, it is also the less sensitive to the baryon-to-photon fraction. Current $Y_P$ determinations have an accuracy of about 0.003, an accuracy that is not high enough for some applications. To improve further the $Y_P$ determinations one must take into account all the sources of uncertainty that affect, down to the 0.0005 level, the precision of $Y_P$ determinations.

General reviews on primordial nucleosynthesis have been presented by Boesgaard and Steigman (1), Olive, Steigman, & Walker (2), and Steigman (3), and a historical note on the primordial helium abundance has been presented by Peimbert and Torres-Peimbert (4).

## 2. OBSERVATIONS IN THE SIXTIES

The lack of precision in the determination of the helium abundance and the lack of knowledge on the processes of gravitational settling of helium and heavy elements in stellar photospheres had led to two radically different ideas to explain the observed values of $Y$(5, 6): (a) galaxies were formed from a gas made of hydrogen without helium, and the relatively high abundance of helium that is observed in young stars and in the interstellar gas was produced by normal stars during the whole life of the galaxies and by supermassive stars at the beginning of galaxy formation, or (b) galaxies formed with an appreciable amount of helium, probably produced during the initial stages of expansion of the universe, as predicted by the Big Bang Theory (BBT). The first possibility implies that the value of $Y$ for very old objects has to be considerably smaller than 0.2, while the second possibility implies values of $Y$ in the 0.2 to 0.3 range for all the old objects.

To decide between these two possibilities it was important to try to find out if there were significant differences among the oldest objects, and in particular if the minimum value of *Y* for old objects was around 0.27 or close to zero, because it was expected that the *Y* values for the oldest objects were closer to the primordial value $Y_P$.

In what follows I present some determinations representative of the *Y* values derived in the sixties.

The values of *Y* obtained from H II regions, the Sun, planetary nebulae from the galactic disk, and young stars of type B were of 0.32, 0.32, 0.36, and 0.37 respectively (7,8). These abundances correspond to relatively young objects and had errors in the 0.05 to 0.10 range.

H II regions are clouds of gas ionized by stars recently made in these clouds. The H II regions are very young with typical ages of one to three million years. The Sun has an age of about 4.5 billion years. Planetary nebulae are clouds of ionized gas ejected by intermediate mass stars when they evolve from the stage of red giants to the stage of white dwarfs; galactic disk planetary nebulae have typical ages of about one billion years, while halo planetary nebulae have typical ages of about ten billion years.

O'Dell, Peimbert, & Kinman, (9), found that $Y = 0.42 \pm 0.10$ for a planetary nebula in the stellar cluster M15, this cluster is very old, its content of heavy elements is about two orders of magnitude smaller than solar, it is located in the halo of the galaxy and it was formed from gas with a chemical composition very close to the primordial one. Since the *Y* value for this planetary nebula was similar to the values of young planetary nebulae in the disk of the galaxy we concluded that most of the helium was formed before the galaxy was formed, result that favored the BBT.

Sargent and Searle (10) and Greenstein and Munch (11) found values of *Y* smaller than 0.03 in the photosphere of several very old stars, apparently in contradiction with the BBT. Later on Greenstein, Truran, and Cameron (12) suggested that the helium atoms in the photosphere diffused towards the center of the stars due to differential diffusion relative to the hydrogen atoms. On the other hand these stars show overabundances of some heavy elements in their surface, apparently against the idea of separation by gravitational settling of the different elements (5, 6).

## 3. BIG BANG PREDICTIONS IN THE SIXTIES

I will not mention here work done before the sixties. For those interested in the BBN predictions made before the sixties consult the papers by Hoyle and Tayler (13), Peebles (14), and Wagoner, Fowler, & Hoyle (15).

Hoyle & Tayler (13) based on computations of BBN with two families of neutrinos found that $Y_P = 0.36 \pm 0.04$. They mentioned that this value of $Y_P$ probably is a lower limit of the true value because the number of neutrino families might be larger than two. They stated that if it is established empirically that the value of *Y* is considerably smaller than 0.36 in any astronomical object in which there has not been separation of hydrogen and helium by gravitational settling, it can be concluded that the universe did not originate during a big bang.

The discovery of the fossil or cosmic microwave background radiation (CMB) in 1965 gave new impetus and support to the BBT and led Peebles (14) to produce a new set of computations of the nuclear reactions that take place during the big bang with a higher precision than before. He assumed two neutrino families and a present temperature for the CMB of 3 degrees Kelvin and found that for a wide set of values for the baryonic density of the universe $Y_P$ lies in the 0.26 to 0.28 range.

Afterwards Wagoner, Fowler, & Hoyle (15) produced new computations similar to those by Peebles (14). They found that the $Y_P$ value predicted by BBN was in the 0.2 to 0.3 range. They also made computations for supermassive stars with more than 1000 solar masses that, if existed, could produce $Y$ values even higher. They concluded that if it was confirmed that young stars and planetary nebulae have $Y$ values of the order of 0.4, this could be an indication that the helium was produced by supermassive stars, and alternatively if the value of $Y$ is close to 0.27 then it would be an evidence in favor of a universal fireball, (in other words the big bang).

## 4. $Y_P$ DETERMINATIONS FROM OXYGEN POOR H II REGIONS

In the seventies the differences among the $Y$ values for old objects determined observationally were reduced drastically, mainly due to the higher accuracy of the determinations, and it was concluded that the most direct way to determine $Y_P$ is by observing extragalactic H II regions with very small amounts of heavy elements. Figures 1 and 2 show two very popular extragalactic H II regions that have been used to determine $Y_P$.

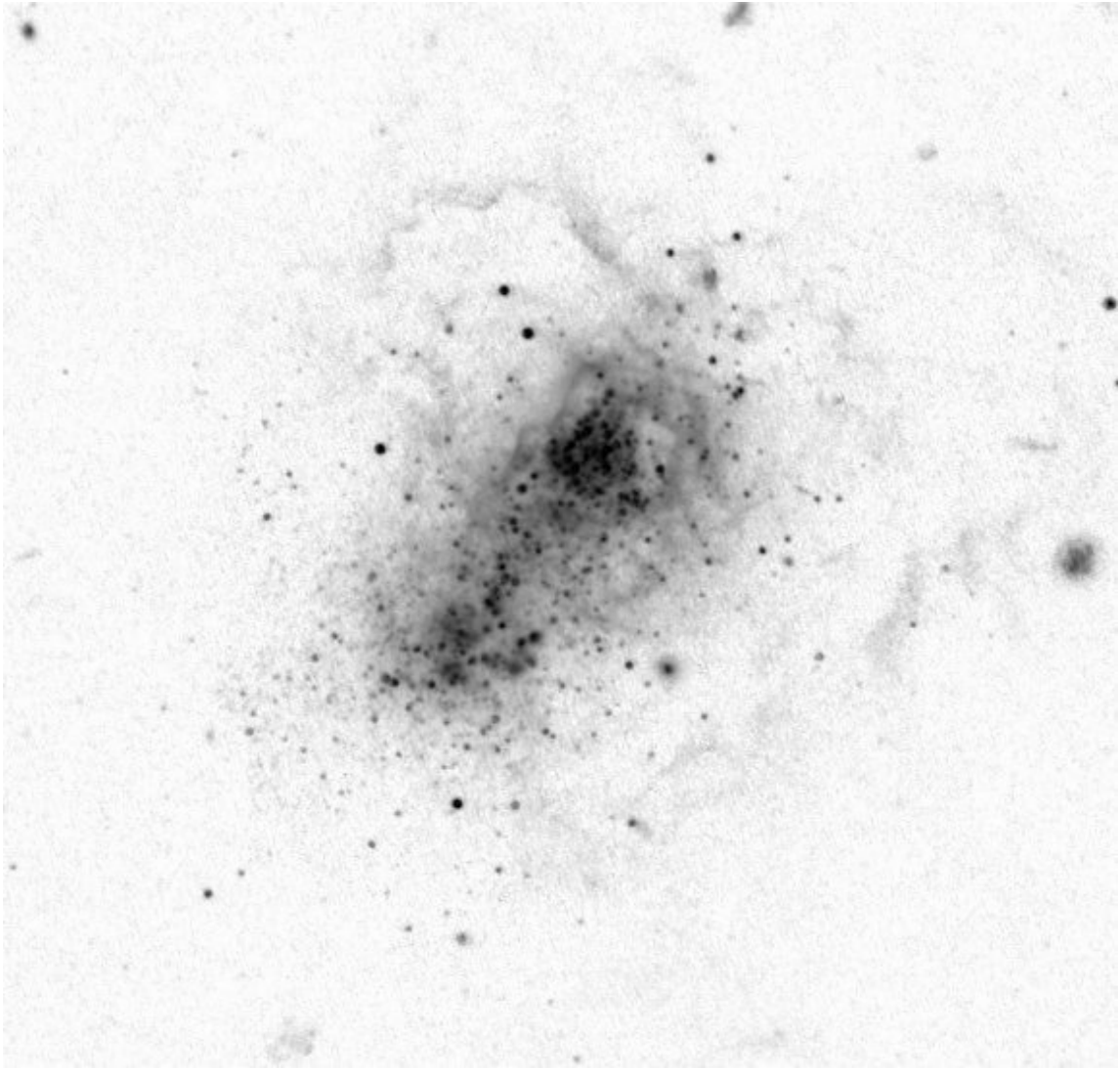

Figure 1. Star formation region in the irregular galaxy I Zwicky 18 probably the most metal poor galaxy known to date with relatively bright regions of star formation. This galaxy has most of its baryonic mass in the form of gas that is almost uncontaminated by the products of stellar evolution. By obtaining the chemical composition of galaxies like this it is possible to find the initial chemical abundances with which galaxies were formed, this composition is of approximately 25% by helium and 75% of hydrogen by unit mass. Image from NASA's Hubble Space Telescope. Courtesy of Y. Izotov (Main Astronomical Observatory, Kyiv, Ucrania) and T. Thuan (University of Virginia). Image from NASA'S Hubble Space Telescope.

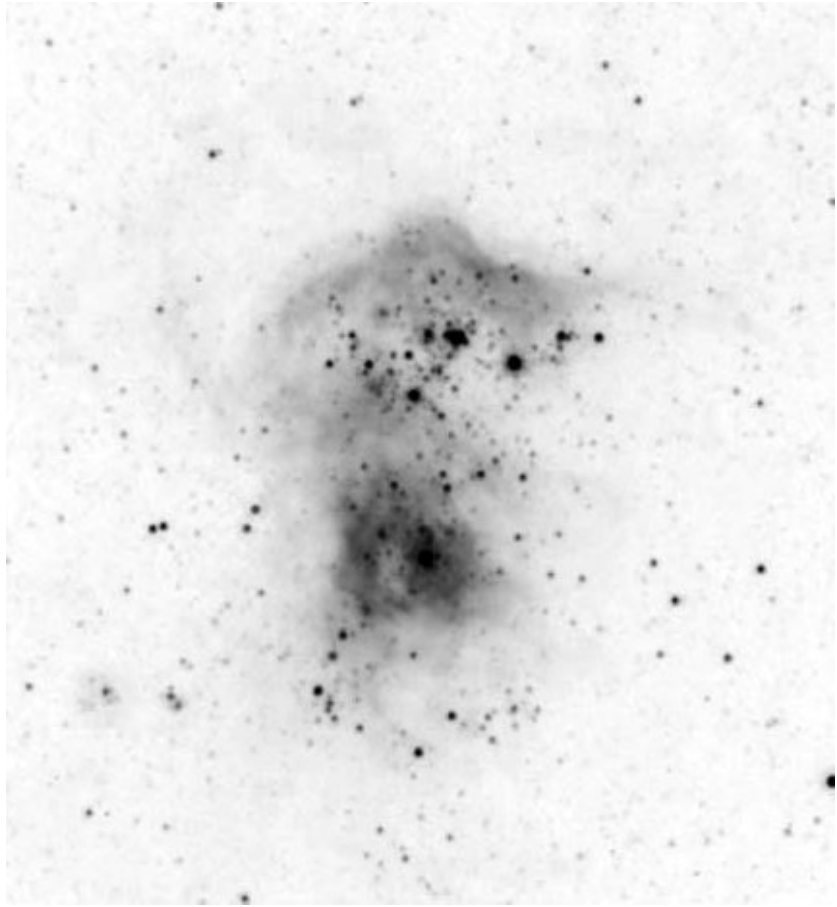

Figure 2. H II region NGC 2363 located in the irregular galaxy NGC 2366. This is one of the brightest extragalactic H II regions with a relatively low content of heavy elements. This object has been studied by many different groups to determine $Y_P$. Courtesy of L. Drissen, J. Roy, and C. Robert (University of Laval), and Y. Dutil (Canadian French Hawaiian Telescope). Image from NASA'S Hubble Space Telescope.

Searle & Sargent (16) analyzed the evidence on the possibility that the $Y$ values in the stellar photospheres of the stars with very small amounts of helium were representative of the whole star, and consequently that they had cosmological implications, and concluded that this was not the case. Searle & Sargent (17) obtained for $Y$ values of $0.29 \pm 0.05$ and $0.25 \pm 0.05$ respectively, for I Zwicky 18 and II Zwicky 40 two irregular galaxies with: a low content of heavy elements, a large fraction of gaseous mass, and a low fraction of stellar mass; these properties imply that the interstellar gas of these galaxies has not been contaminated appreciably by gas enriched in heavy elements and helium ejected to the interstellar medium by stellar winds, planetary nebulae formation, and the explosion of supernovae. These $Y$ abundances turned out to be similar to those of H II regions of the solar vicinity, result that supports the idea that the galaxies had been formed with a value of $Y_P$ in the 0.2 to 0.3 range. In 1973 Peimbert (18) found that that the planetary nebula in the stellar cluster M15 has a $Y$ value of $0.29 \pm 0.03$; the higher precision than that presented in (9) is due mainly to observations of higher quality.

Peimbert & Torres Peimbert (19,20) and Lequeux et al. (21) found differences between the helium abundances of the H II regions in the Magellanic Clouds (two irregular galaxies very close to our galaxy) and the Orion nebula (an H II region of our galaxy) larger than the observational errors. These differences imply that the stars have ejected part of the helium present in the interstellar medium. Based on observations of a group of extragalactic H II regions and on a linear relationship between $Y$ and $Z$ extrapolated to $Z = 0$, Lequeux et al. (21) found that $Y_P = 0.233 \pm 0.014$.

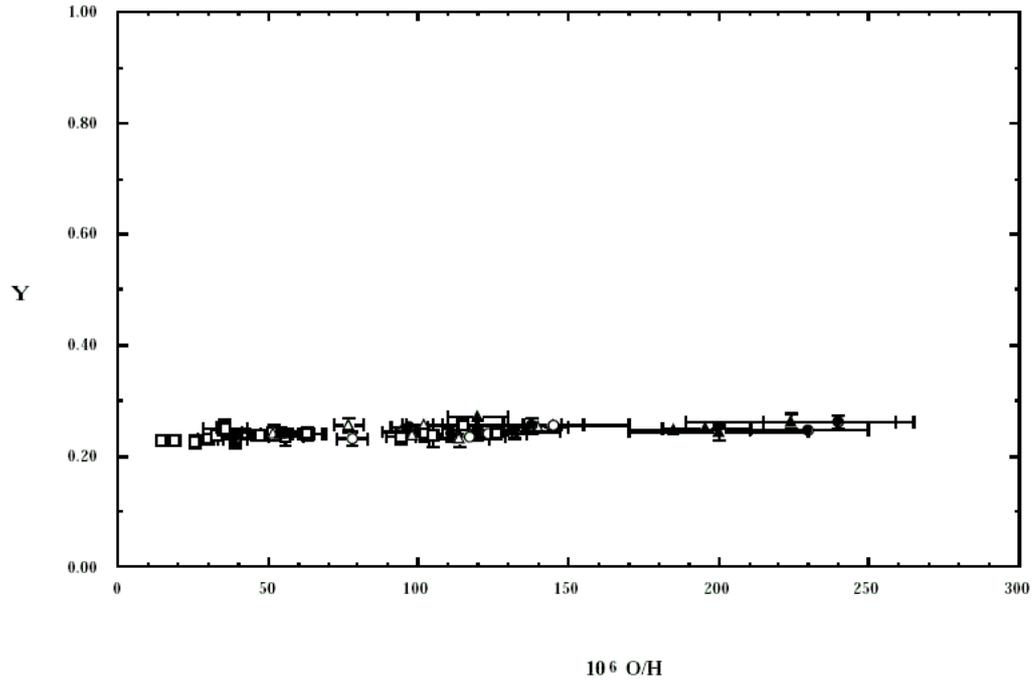

Figure 3. The $^4$He mass fraction Y versus the O/H value, abundances derived from observations of low-metallicity extragalactic H II regions.

Many determinations of the $Y_P$ value have been derived since the seventies. In Figure 3 a compilation of $Y$ determinations for $O$-poor extragalactic H II regions carried out by Steigman (22) is presented. In Figure 4 a small fraction of a recent spectrogram of 30 Doradus, the brightest extragalactic H II in the Large Magellanic Cloud, is presented.

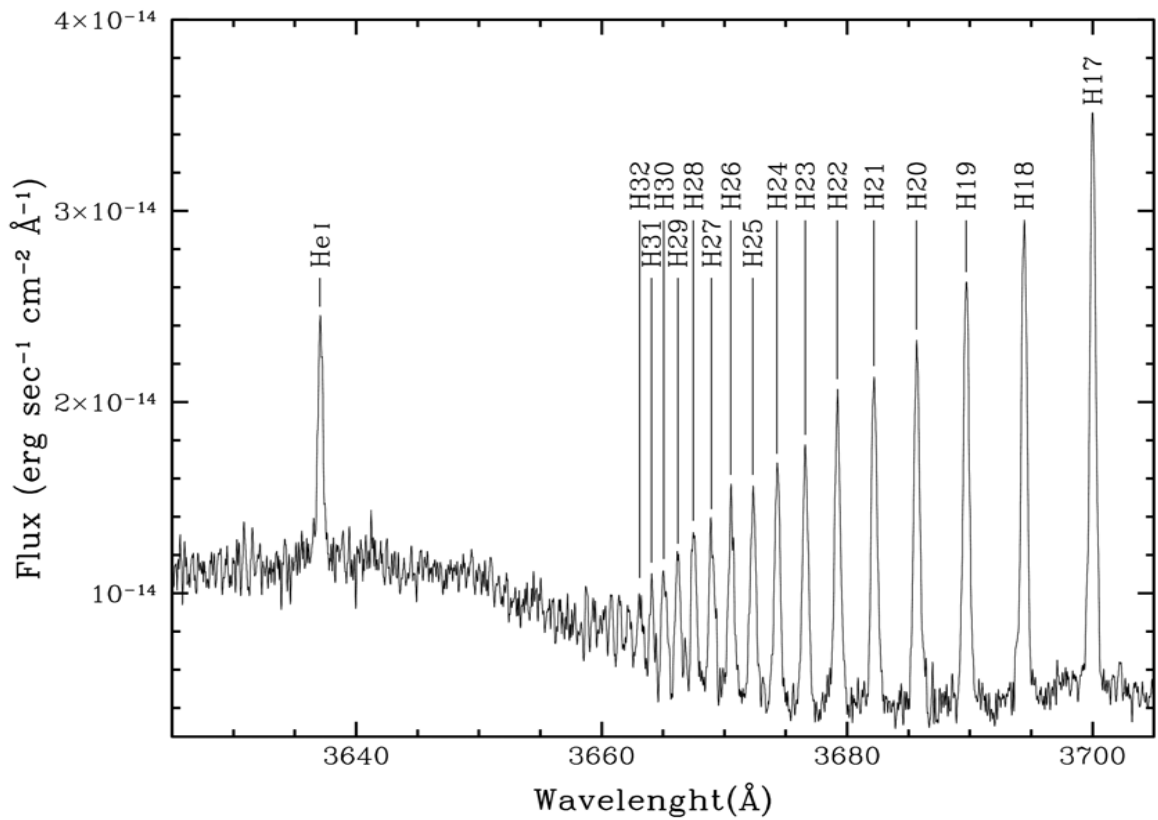

Figure 4. Small fraction of a high resolution spectrogram of the H II region 30 Doradus in the Large Magellanic Cloud. The spectrogram was taken with the Echelle spectrograph of the Very Large Telescope (VLT) in Paranal Chile (23). Note that it is possible to detect hydrogen transitions up to Balmer 32. The emission line at 3637 A is a relatively faint He I line.

TABLE 1
PRIMORDIAL HELIUM DETERMINATIONS

| Year | $Y_P$ | Statistical Error | Systematic Error | References |
|------|-------|-------------------|------------------|------------|
| 1992 | 0.2280 | ± 0.0050 | ± 0.0200 | (24), this paper |
| 2000 | 0.2345 | ± 0.0026 | ± 0.0120 | (25), this paper |
| 2003 | 0.2391 | ± 0.0020 | ± 0.0075 | (26), this paper |
| 2004 | 0.2421 | ± 0.0021 | ± 0.0085 | (27), this paper |
| 2007 | 0.2516 | ± 0.0011 | ± 0.0045 | (28), this paper |
| 2007 | 0.2477 | ± 0.0018 | ± 0.0023 | (29) |

In Table 1 I present some of the best $Y_P$ determinations of the last few years (24-29). Table 1 includes the errors presented by the authors (mainly statistical), plus the systematic errors estimated in this review and not considered by the authors, with the exception of the $Y_P$ determination by Peimbert, Luridiana, & Peimbert (29) that does include estimates of the statistical and systematic errors. The differences among all the determinations of $Y_P$ are mainly due to systematic errors.

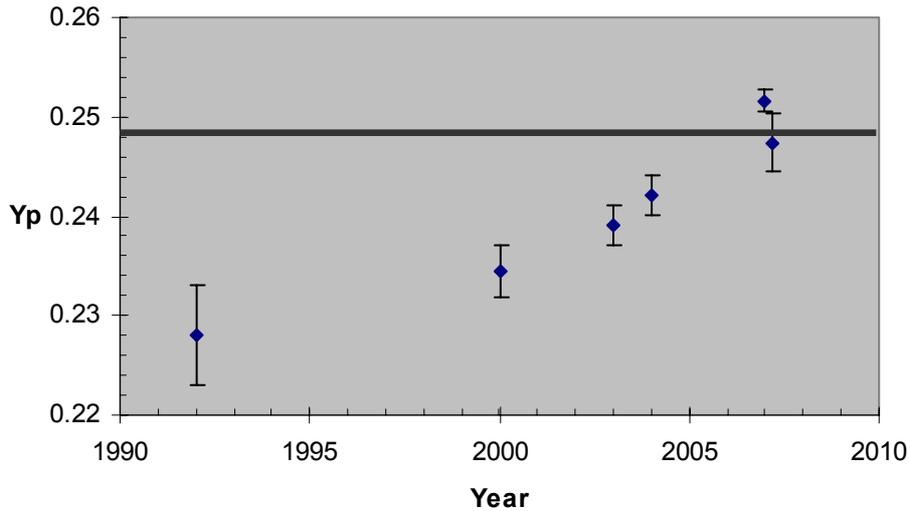

Figure 5. $Y_P$ determinations derived from H II regions with a small fraction of heavy elements, usually called metal poor or extremely metal-poor H II regions. The $Y_P$ values presented are those of Table 1 and only include the statistical errors presented by the authors with the exception of the $Y_P$ value by Peimbert et al. (29) that amounts to 0.2479. ± 0.0029 that also includes an estimate of the systematic errors. The horizontal line is the $Y_P$ value estimated from the baryon to photon density of the universe estimated by Dunkley et al. (30) from the *Wilkinson Microwave Anisotropy Probe (WMAP)* observations under the assumption of SBBN and adopting a lifetime for the neutron, $\tau_n$, of 885 sec determined by Arzumanov et al. (31).

The first five determinations presented in Table 1 include one or more of the following systematic errors: (a) The observed spectra that shows the He I lines in emission produced in the nebula also includes the same underlying He I lines in absorption produced by the ionizing stars, even if in the determination of the $Y$ abundances of nearby H II regions like NGC 346 in the Small Magellanic Cloud it is possible to avoid the ionizing stars in the observing slit, a good fraction of the observed nebular continuum is due to dust scattered light that shows the stellar He I lines in absorption, this systematic problem is now well studied based on the stellar population synthesis models developed by González Delgado et al. (32, 33) that predict the stellar continuum expected as a function of time, if the correction for underlying He I absorption is not considered with spectra of low resolution it is possible to underestimate $Y_P$ by as much as 0.0080 by unit mass; (b) the new He I recombination coefficients by Porter et al. (34, 35) increase $Y_P$ by about 0.0040 by unit mass relative to the values computed by Smits (36) and Benjamin, Skillman, & Smits (37); (c) in high-temperature H II regions, the hydrogen Balmer lines can be excited by collisions of neutral atoms with free electrons, this effect is generally small, usually contributing less than a few percent of the intensity of Hα and Hβ, that for most applications can be neglected, however, collisional enhancement must be taken into account in the determination of $Y_P$ for at least two reasons, first, for this task the helium abundance by unit mass, $Y$, in individual H II regions must be known with the highest attainable precision, second, the objects where collisions are more important are those with the highest electron temperatures and, hence, the lowest metallicities; consequently, they bear a major weight in the extrapolation to $Z = 0$ of the relation between $Y$ and $Z$, moreover the recent results by Anderson et al. (38, 39) indicate that the collisional excitation coefficients are higher than previously thought, the correction for collisional excitation of the Balmer lines depends on the particular sample of H II regions used to determine $Y_P$ but usually produces an increase in $Y_P$ from 0.0030 to 0.0070 mass units, therefore if this effect is neglected $Y_P$ is underestimated (26); (d) the $Y$ value of individual H II regions decreases when the temperature structure of the H II region is taken into account, since in the self-consistent solutions for all the observed He I line intensities, the lower $T$(He II) values imply higher densities, the higher densities produce a higher collisional contribution to the He I intensities, and consequently lower helium abundances, this effect decreases the $Y_P$ determination by about 0.0040, this effect will be discussed further in the next two sections.

As discussed above most of the systematic effects not taken into account in the past produce an underestimation of the $Y_P$ value. The increase in the $Y_P$ determination seen in Figure 5 is due to a decrease in the systematic errors as shown in Table 1.

5. ERROR BUDGET OF THE MOST RECENT $Y_P$ DETERMINATION

In Table 2 I present the error budget by Peimbert, Luridiana, & Peimbert (29) on the $Y_P$ determination that includes 13 main sources of error. In this section I will discuss only the four main sources of error; a thorough discussion of all the sources of error is presented in the original paper.

TABLE 2

ERROR BUDGET FOR THE $Y_P$ DETERMINATION BY PEIMBERT, LURIDIANA, & PEIMBERT (29)

| Source | Estimated Error |
|---|---|
| Collisional excitation of the H I lines | $\pm 0.0015$[a] |
| Temperature structure | $\pm 0.0010$[b] |
| $O(\Delta Y/\Delta O)$ correction | $\pm 0.0010$[a] |
| Recombination coefficients of the He I lines | $\pm 0.0010$[a] |
| Collisional excitation of the He I lines | $\pm 0.0007$[b] |
| Underlying absorption in the He I lines | $\pm 0.0007$[b] |
| Reddening correction | $\pm 0.0007$[a] |
| Recombination coefficients of the H I lines | $\pm 0.0005$[a] |
| Underlying absorption in the H I lines | $\pm 0.0005$[b] |
| Helium ionization correction factor | $\pm 0.0005$[b] |
| Density structure | $\pm 0.0005$[b] |
| Optical depth of the He I triplet lines | $\pm 0.0005$[b] |
| He I and H I line intensities | $\pm 0.0005$[b] |

[a]Systematic error.
[b]Statistical error.

5.1. *Collisional Excitation of the Hydrogen Balmer Lines*

The most important source of error is the collisional excitation of Balmer lines. Neutral hydrogen atoms in excited states may form not only by the usual process of $H^+$ recombination, but also through collisions of neutral hydrogen with electrons. The recombination cascade that follows such excitations contributes to the observed intensity of Balmer lines, mimicking a larger relative hydrogen abundance and, hence, a smaller helium abundance. To estimate this contribution it is necessary to have a tailored photoionization model for each object that fits properly the temperature structure. The contribution to the Balmer line intensities depends strongly on the temperature: therefore this effect, and consequently the associated error in its estimate, increases for H II regions of lower metallicity and consequently higher temperature.

This uncertainty has in turn three independent sources: the theoretical uncertainty on the collisional excitation coefficients, $\Omega$s, that not all the $\Omega$s for higher n values have been computed, and the uncertainty on the physical conditions of the gas at which collisions occur.

The last of these factors is probably the most severe. The collisional excitation of the Balmer lines is stronger in the hotter zones of the H II regions (which are predicted to be the innermost in this metallicity range) and in the less ionized zones (which are the outermost). Since the fraction of neutral hydrogen in a nebula varies over a much wider range than the Boltzmann factor, which is the main temperature dependence of collisions, the contribution of the outer zones dominates, so that $T(O^+)$ is the most appropriate temperature to characterize collisions when trying to model them.

It is clear that only detailed observations and tailored modeling will allow us to reduce the uncertainty introduced by the collisional excitation of the Balmer lines, particularly in the extreme cases; but a better choice might be to avoid critical H II regions altogether, by preferentially selecting those H II regions in which collisions are known in advance to play only a minor role, i.e. objects with moderate temperatures. Since the temperature in H II regions is mostly determined by metallicity, this amounts to saying that metal-poor objects are more adequate than extremely metal-poor objects in $Y_P$ determinations. This conclusion runs counter the common wisdom that the best candidates for primordial helium determinations are extremely metal-poor objects ($Z \leq 0.0005$), because they permit to minimize the error introduced by the extrapolation of the ($Y$, $O$) relation to O/H = 0.00: although this is true, a larger uncertainty on $Y_P$ is introduced by collisions than the one introduced by most of the other sources, including the slope $\Delta Y/\Delta O$, so the observational efforts should be better directed at metal-poor objects.

Further disadvantages of extremely metal-poor objects in the quest for primordial helium are that their number is very small and that their H II regions are relatively faint. These disadvantages, together with the uncertainty on collisions discussed above, largely outweigh the advantage implied by the smaller error introduced by the extrapolation of the ($Y,O$) relation to zero metallicity. Therefore, we are led to the strong conclusion that not so extremely metal poor objects, like those in the $0.0005 \leq Z \leq 0.001$ range, are more appropriate for the determination of $Y_P$: in this range of metallicity it is possible to find a large set of objects with bright H II regions, which will improve the quality and number of emission lines available for the determination of physical conditions; furthermore, the temperatures of these objects are smaller than those of extremely metal-poor objects and consequently the correction due to the collisional excitation of the H I lines is also smaller.

This conclusion leads us to another critical point in the approach to $Y_P$. Olive & Skillman (40) have noted that the uncertainty on $Y_P$ is dominated by systematic errors. As long as this is the case, it is preferable to analyze a few objects in depth and try to correct for the systematics than to perform a more shallow analysis of a larger sample, since this last method can reduce the statistical uncertainties but not the systematic errors. Hence we strongly support the methodological choice of understanding as well as possible the details of the objects in a small sample, before directing our efforts towards extending the sample. It is only by means of this approach that systematical errors, such as those arising from the temperature structure of H II regions, can be gradually understood, corrected for, and eventually transformed into statistical uncertainties.

5.2. *Temperature structure of H II regions*

The second most important source of error is the temperature structure. Most determinations neglect the possible presence of temperature variations across the H II region structure and assume that the electron temperature derived from the ratio of the λλ 4363 A, 5007 A [O III] lines, $T$(O III), is representative of the zone where He I recombination lines form. However, other temperature determinations based on different diagnostics yield lower values; furthermore, photoionization models do not predict the high $T$(O III) values observed. These results indicate that temperature variations are indeed present in H II regions, and this result should be included in the $Y$ determination (41). The best procedure to take into account the temperature structure is to self-consistently determine the temperature where the He I lines originate, $T$(He II), from a set of He I lines by means of the maximum likelihood method. The $Y$ abundances derived from $T$(He II) are typically lower by about 0.0030 to 0.0050 than those derived from $T$(O III). The difference between both temperatures does not have a significant trend with the metallicity of the H II region, hence the systematic error introduced by the use of $T$(O III) in the $Y$ determination is similar for objects with different metallicities.The error quoted under "Temperature Structure" in Table 2 is the residual error due to the uncertainty in the $T$(He II) determinations of the dataset by Peimbert, Luridiana, & Peimbert (29).

5.3. *Extrapolation of the Y Determinations to the Value of $Y_P$, or the $O(\Delta Y/\Delta O)$ Correction*

To determine the $Y_P$ value from a set of metal poor H II regions it is necessary to estimate the fraction of helium, present in the interstellar medium of the galaxy where each H II region is located, produced by galactic chemical evolution. We will assume that

$$Y_P = Y - O(\Delta Y/\Delta O), \qquad (1)$$

where $O$ and $Y$ are the abundances of the observed region. From chemical evolution models of different galaxies it is found that $\Delta Y/\Delta O$ depends on the initial mass function (IMF), the star formation rate, the age, and the $O$ value of the galaxy in question. Peimbert et al. (42) have found that $\Delta Y/\Delta O$ is well fitted by a constant value for objects with the same IMF, the same age, and an $O$ abundance smaller than $4 \times 10^{-3}$ (42). Consequently for metal poor galaxies, that by definition have $O$ abundances smaller than $4 \times 10^{-3}$, it is customary to adopt a constant value for $\Delta Y/\Delta O$.

To obtain an accurate $Y_P$ value, a reliable determination of $\Delta Y/\Delta O$ for $O$-poor objects is needed. The $\Delta Y/\Delta O$ value derived by Peimbert, Peimbert, & Ruiz (25) from observational results and models of chemical evolution of galaxies amounts to $3.5 \pm 0.9$. More recent results are those by Peimbert (23) who finds $2.93 \pm 0.85$ from observations of 30 Dor and NGC 346, and by Izotov & Thuan (27) who, from the observations of 82 H II regions, find $\Delta Y/\Delta O = 4.3 \pm 0.7$. Peimbert, Luridiana, & Peimbert (29) have recomputed the value by Izotov & Thuan considering two systematic effects not considered by them: the fraction of oxygen trapped in dust grains, which we estimate to be 10% for objects of low metallicity, and the increase in the $O$ abundances due to the presence of temperature fluctuations, which for this type of H II regions we estimate to be about 0.08 dex (43). From these considerations we obtain for the Izotov & Thuan sample a $\Delta Y/\Delta O = 3.2 \pm 0.7$.

On the other hand Peimbert et al. (42) from chemical evolution models with different histories of galactic inflows and outflows for objects with $O < 4 \times 10^{-3}$ find that $2.4 < \Delta Y/\Delta O < 4.0$. From the theoretical and observational results we have adopted a value of $\Delta Y/\Delta O = 3.3 \pm 0.7$, that we have used with the $Y$ and $O$ determinations from each object to obtain the set of $Y_P$ values, whose average value is presented in the last line of Table 1 and the estimated error due to this extrapolation is presented in Table 2.

### 5.4. *Recombination Coefficients of the Helium I Lines*

The fourth most important source of error is the uncertainty on the recombination coefficients of the He I lines. We have estimated the error in the $Y_P$ determination due to the adopted effective recombination coefficients based on the confidence in the He I line emissivities presented by Bauman et al. (44). The lines used to determine the helium abundances are $\lambda\lambda$ 3820(A), 3889(B), 4026(AA), 4387(A), 4471(A), 4921(A), 5876(A), 6678(A), 7065(A), and 7281(A), where the letter indicates the confidence: (AA) better than 0.1%, (A) in the 0.1 to 1% range, and (B) in the 1 to 5% range. From these values we estimate a systematic error due to the computed emissivities of about 0.0010 in $Y$. According to Porter, Ferland, & MacAdam (35), the error introduced in the emissivities by interpolating the equations provided by them in temperature is smaller than 0.03%, which translates into an error in $Y_P$ considerably smaller than 0.0001.

## 6. COMPARISON OF THE LAST TWO $Y_P$ DETERMINATIONS

The procedures used by Izotov, Thuan, & Stasińska (28) and Peimbert, Luridiana, & Peimbert (29) to determine $Y_P$ are very different and it is not easy to make a detailed comparison of all the steps carried out by both groups. I consider that the error presented by Izotov et al. is a lower limit to the total error because it does not include estimates of possible systematic errors. The difference in the central $Y_P$ values derived by both groups is mainly due to the treatment of the temperature structure of the H II regions. Izotov et al. adopt temperature variations that are smaller than those used by Peimbert et al. To be more specific we can define the temperature structure of the H II regions by means of an average temperature, $T_0$, and a mean square temperature fluctuation, $t^2$; these quantities are given by (45):

$$T_0 = \frac{\int T_e(\vec{r}) N_e(\vec{r}) N_i(\vec{r}) dV}{\int N_e(\vec{r}) N_i(\vec{r}) dV}, \quad (2)$$

and

$$t^2 = \frac{\int [T_e(\vec{r}) - T_0]^2 N_e(\vec{r}) N_i(\vec{r}) dV}{T_0^2 \int N_e(\vec{r}) N_i(\vec{r}) dV}, \quad (3)$$

where $N_e$ and $N_i$ are the electron and the ion densities of the observed emission line and $V$ is the observed volume.

The value of $t^2$ derived by Izotov et al. (28) for their sample is about 0.01; while Peimbert et al. (29) obtain a $t^2$ of about 0.025. From the computations by Peimbert et al. and adopting $t^2 = 0.000$ we obtain $Y_P = 0.2523 \pm 0.0027$, and for $t^2 = 0.01$ we obtain $Y_P = 0.2505$.

The small value of $t^2$ derived by Izotov et al. (28) is due to the parameter space used in their Monte Carlo computation where they permitted $T$(He II) to vary from 0.95 to 1.0 times the $T$(O III) value, which yields a $t^2$ of about 0.01. By allowing their $T$(He II) to vary from 0.85 to 1.0 times the $T$(O III) value their $t^2$ result would have become higher. This can be seen from their Table 5 where 71 of the 93 objects corresponded to the lowest $T$(He II) allowed by the permitted parameter space of the Monte Carlo computation. A higher $t^2$ value for the Izotov et al. (28) sample produces a lower $Y_P$, reducing the difference with the Peimbert et al. (29) $Y_P$ value.

Two other systematic problems with the Izotov et al. (28) determination related with the temperature structure are: (a) that to compute the $N(O^{++})$ abundance they adopted the $T$(O III) value that is equivalent to adopt $t^2 = 0.00$, and (b) to compute the once ionized oxygen and nitrogen abundances, $N(O^+)$ and $N(N^+)$, they adopted the $T$(O III) value, but according to photoionization models and observations of $O$-poor objects, the temperature in the $O^+$ regions, $T(O^+)$, is considerably smaller than in the $O^{++}$ regions, typically by about 2000 K for objects with $T$(O III) ~ 16000 K and reaching 4000 K for the metal poorest H II regions, see for example the work by Peimbert, Peimbert, & Luridiana (41) and by Stasińska (46).

Another argument that possibly implies that the temperature structure was not properly considered by Izotov et al. (28) is the $Y_P$ value obtained by them from a linear regression of $Y$ and N/H to N/H equal to zero, $Y_P(N)$, that amounts to $0.2532 \pm 0.0009$. They argue that the higher $Y_P(N)$ value with respect to the $Y_P(O)$ value is due to a faster increase of $N$ relative to the increase of $O$, it can be shown that in this case the $Y_P(N)$ should be smaller than the $Y_P(O)$, not higher as they find. Two other systematic effects that might increase their error in the $Y_P(N)$ determination are the adoption of $T$(O III) for the $N^+$ and $O^+$ regions mentioned above, and the assumption that the $N^+$ and $O^+$ regions are coincident, which is not always the case (47).

## 7. INITIAL CONDITIONS FOR STELLAR EVOLUTION MODELS AND THE CHEMICAL EVOLUTION OF GALAXIES

*7.1 Metal poor galaxies*

As discussed in section 5.3 from observations it has been obtained that $\Delta Y/\Delta O = 3.3 \pm 0.7$ for objects with an $O$ abundance smaller than $4 \times 10^{-3}$. It is also possible from models of chemical evolution of galaxies to determine $\Delta Y/\Delta O$, this ratio depends on: (a) an initial stellar mass function, IMF, (b) a set of stellar evolution models, (c), a star formation rate, SFR, and (d) the presence of $O$ rich galactic outflows. The $\Delta Y/\Delta O$ ratio does not depend on: (a) the IMF for masses smaller than 0.85 solar masses, (b) the presence of non baryonic matter, (c) the presence of inflows of primordial material, because for this material by definition $\Delta Y$ and $\Delta O$ are equal to zero, and (d) the presence of outflows of well mixed material.

The helium yield is defined by

$$y(Y) = M(Y)/M^*, \qquad (4)$$

where $M(Y)$ is the mass that a generation of stars ejects as newly formed helium by them to the interstellar medium and $M^*$ is the mass of the same generation that remains locked into stellar remnants and long lived stars, where we include the low mass end of the IMF which might comprise objects that do not become stars. Similarly the oxygen yield is defined by

$$y(O) = M(O)/M^*. \qquad (5)$$

Carigi et al. (48, 49, 50) have ruled out the presence of significant outflows of $O$-rich material based on the low $C/O$ values observed in irregular galaxies, with an $O$ abundance smaller than $4 \times 10^{-3}$. Therefore it can be shown that for these objects $\Delta Y/\Delta O = y(Y)/y(O)$.

Peimbert et al. (42) computed a set of galactic chemical evolution models with a constant SFR, an age of 13 billion years, the stellar yields adopted by Carigi et al. (50), the IMF by Kroupa et al. (51), and a maximum mass for the IMF of 80 solar masses and found that $y(Y)/y(O) = 3.3$ for a final $O$ abundance smaller than $4 \times 10^{-3}$.

Carigi et al. (50) and Peimbert et al. (42) computed two sets of models varying one or two of the adopted assumptions with relation to the set mentioned in the previous paragraph: (a) they reduced the maximum stellar mass of the IMF to 60 solar masses and found that $y(Y)/y(O) = 4.0$, since the O production is due to stars with more than 8 solar masses the reduction in their number is mainly responsible for the increase in the $y(Y)/y(O)$, and (b) by adopting a set of models with an age of one billion years and a SFR 13 times higher than the first set they found that $y(Y)/y(O) = 2.6$, the decrease in this value is due to the delay in the helium enrichment of the interstellar medium by intermediate and low mass stars, those with less than 8 solar masses at birth. Based in these three sets of models it is concluded that $\Delta Y/\Delta O = 3.3 \pm 0.7$ is a representative value for models of metal poor galaxies.

To determine the $Y_P$ value from $O$-poor extragalactic H II regions, those with $O < 4 \times 10^{-3}$, we recommend to use $\Delta Y/\Delta O = 3.3 \pm 0.7$. This recommendation is based on the observations discussed in section 5.3, and the galactic chemical evolution models just discussed.

*7.2 Our Galaxy*

Carigi and Peimbert (52) have computed chemical evolution models for the disk of our Galaxy at four galactocentric distances (see Figure 6). The chemical evolution models reproduce the O/H present day gradient in the interstellar medium of the Galactic disk, the O/H ratio diminishes with galactocentric distance. The models assume: (a) a galactic inside-outside formation scenario with gaseous infalls from the intergalactic medium, but without any types of gas outflows to the intergalactic medium, (b) the stellar initial mass function by Kroupa et al. (51) with two maximum stellar masses, 60 and 80 solar masses, and (c) two sets of stellar yields described below. For stars with $O$ abundances higher than $4 \times 10^{-3}$ there are two sets of stellar yields available to compute the chemical evolution models. Since the main difference between these sets is the mass loss rate assumed for massive stars they have called them the high-wind yields (HWY) and the low-wind yields (LWY). The HWY set is the one considered in the best model (model 1) by Carigi et al. (53), this set includes the yields for massive stars by Maeder (54), those in the 8 to 80 solar mass range, for $Z = 0.02$; while the LWY set differs from the HWY set in that it includes the massive star yields for $Z = 0.02$ by Hirschi et al. (55) instead of those by Maeder (54).

Carigi and Peimbert (52) present the following two equations to estimate the initial helium abundance with which stars have formed during the history of the Galactic disk: (a): for $O \leq 4.3 \times 10^{-3}$

$$Y = Y_P + (3.3 \pm 0.7)O, \qquad (6)$$

and for $4.3 \times 10^{-3} \leq O \leq 9 \times 10^{-3}$, using the HWY set

$$Y = Y_P + (3.3 \pm 0.7)O + (0.016 \pm 0.003)\left(\frac{O}{4.3 \times 10^{-3}} - 1\right)^2. \qquad (7)$$

With the use of the LWY set instead of the HWY set the quadratic term in equation (7) becomes negligible.

To test the galactic chemical evolution of helium it is necessary to determine the $Y$ and $O$ abundances in H II regions of the Galactic disk. Unfortunately $O$-rich H II regions, like those of the Galaxy, have only a handful of early type stars and often they are not hot enough to ionize all the helium present in the H II region, the presence of a large fraction of neutral helium makes very difficult to determine the total helium abundance accurately. At present the only exception is the H II region M17, where the fraction of neutral helium is very small and therefore the accuracy with which the $Y$ abundance can be determined is high. M17 is located at a galactocentric distance of 22 thousand light years and shows a higher metallicity than the solar one. The M17 abundances derived by Carigi & Peimbert (52) for $t^2 = 0.000$ amount to $Y = 0.2926 \pm 0.0034$ and $O = 0.00446 \pm 0.00045$; while for $t^2 = 0.036$ amount to $Y = 0.2837 \pm 0.0044$ and $O = 0.00811 \pm 0.00081$ (see Figure 6, where $\Delta Y = Y - Y_P$). The values derived assuming a constant temperature given by $T$(O III) (the $t^2 = 0.000$ case) do not agree with the chemical evolution model, while the values for $t^2 = 0.036$ are in excellent agreement with it. Moreover the model that uses the LWY set is in good agreement with the $Y$ and $O$ values observed in M17 for $t^2 = 0.036$ but not with those for $t^2 = 0.000$. With the present accuracy of the $\Delta Y/\Delta O$ determination of M17 it is not possible to distinguish between the HWY and the LWY models of galactic chemical evolution.

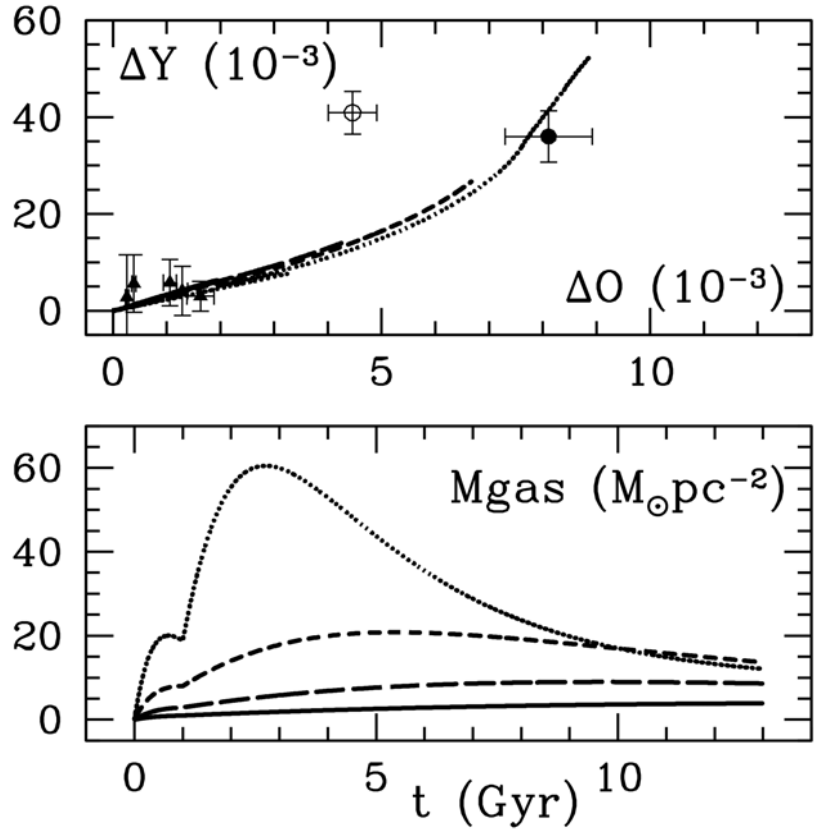

Figure 6. Chemical evolution models for the disk of the Galaxy at four galactocentric distances 52, 39, 26, and 13 thousand light years (continuous, long-dashed, short-dashed, and dotted lines respectively), the Sun is at a galactocentric distance of 26 thousand light years. The upper panel shows the evolution of $Y$ with respect to $O$, and the lower panel the amount of gas in the interstellar medium as a function of time (52). The models in this figure were computed with the HWY set and a maximum stellar mass of 80 solar masses. In the upper panel the $\Delta Y$ and the $\Delta O$ abundances derived from observations for the H II region M17 are presented (filled circle the value for $t^2 = 0.036$, open circle the value for $t^2 = 0.000$), also in this panel the low metallicity H II regions studied by Peimbert, Luridiana, & Peimbert (29) are presented (filled triangles).

The HWY set has been recommended by Carigi et al. (53) to be used in the evolutionary stellar models for massive stars of high metallicity, because only with stellar yields that assume the high mass loss rate for the $Z = 0.02$ models it is possible to reproduce the O/H and C/O gradients across the disk of the Galaxy.

Most of the stars in the universe have photospheric temperatures smaller than 11000 K, and consequently do not show helium lines in absorption, therefore it is very difficult to determine their helium abundance. On the other hand it is possible to determine the oxygen abundance of many stars of different temperatures in our Galaxy.

To produce a set of stellar evolution models with different initial oxygen abundances we recommend the use of equations (6) and (7). If from stellar observations it is possible to estimate the initial $O$ abundance of the star, then equations (6) and (7) can be used to estimate the initial $Y$ abundance of the star, the accuracy of the initial $Y$ abundance will depend on the accuracy of $Y_P$.

## 8. COMPARISON OF THE DIRECTLY DETERMINED $Y_P$ WITH THE $Y_P$ VALUES COMPUTED UNDER THE ASSUMPTION OF STANDARD BBN AND THE OBSERVATIONS OF $D_P$ AND *WMAP*

To compare the $Y_P$ value with the primordial deuterium abundance $D_P$ (usually expressed as $10^5(D/H)_P$) and with the *WMAP* results, we will use the framework of the SBBN. The ratio of baryons to photons multiplied by $10^{10}$, $\eta_{10}$, is given by (57):

$$\eta_{10} = (273.9 \pm 0.3)\Omega_b h^2, \qquad (8)$$

where $\Omega_b$ is the baryon closure parameter, and $h$ is the Hubble parameter. In the range $4 \leq \eta_{10} \leq 8$ (corresponding to $0.2448 \leq Y_P \leq 0.2512$), $Y_P$ is related to $\eta_{10}$ by (56):

$$Y_P = 0.2375 + \eta_{10}/625. \qquad (9)$$

In the same $\eta_{10}$ range, the primordial deuterium abundance is given by (56):

$$10^5(D/H)_P = D_P = 46.5\, \eta_{10}^{-1.6}. \qquad (10)$$

From the $Y_P$ value by Peimbert, Luridiana, & Peimbert (29), the $D_P$ value by O'Meara et al. (58), the $\Omega_b h^2$ value by Dunkley et al. (30), and the previous equations we have produced Table 3. From this table, it follows that within the errors $Y_P$, $D_P$, and the *WMAP* observations are in very good agreement with the predicted SBBN values. Also in Figure 5 I present, under the assumption of the SBBN the $Y_P$ value derived from the *WMAP* observations as a horizontal line.

TABLE 3

COSMOLOGICAL PREDICTIONS BASED ON SBBN AND OBSERVATIONS
FOR $\tau_n = 885.7 \pm 0.8$ sec

| Method | $Y_P$ | $D_P$ | $\eta_{10}$ | $\Omega_b h^2$ |
|---|---|---|---|---|
| $Y_P$ | $0.2477 \pm 0.0029$[a] | $2.78^{+2.28}_{-0.98}$[b] | $5.813 \pm 1.81$[b] | $0.02122 \pm 0.00663$[b] |
| $D_P$ | $0.2476 \pm 0.0006$[b] | $2.82 \pm 0.28$[a] | $5.764 \pm 0.360$[b] | $0.02104 \pm 0.00132$[b] |
| *WMAP* | $0.2484 \pm 0.0003$[b] | $2.49 \pm 0.11$[b] | $6.225 \pm 0.170$[b] | $0.02273 \pm 0.00062$[a] |

[a]Observed value.
[b]Predicted value.

Equations (8), (9), and (10) were derived under the assumption of a neutron lifetime, $\tau_n$, of 885.7 ± 0.8 sec (31). A recent result by Serebrov et al. (59) yielded a $\tau_n$ of 878.5 ± 0.7 ± 0.3 sec. This result would lead to an SBBN $Y_P$ value of 0.2468 for the *WMAP* $\Omega_b h^2$ determined value (57, 60).

I present the cosmological predictions included in Table 4 for the new world average of the neutron lifetime presented by Mathews et al. (60), that amounts to $\tau_n$ = 881.9 ± 1.6 sec. This new world average includes the results by Arzumanov et al. (31) and Serebrov et al. (59).

TABLE 4

COSMOLOGICAL PREDICTIONS BASED ON SBBN AND OBSERVATIONS
FOR $\tau_n$, = 881.9 ±1.6 sec

| Method | $Y_P$ | $D_P$ | $\eta_{10}$ | $\Omega_b h^2$ |
|---|---|---|---|---|
| $Y_P$ | 0.2477 ± 0.0029[a] | $2.32^{+1.78b}_{-0.71}$ | 6.375 ± 1.81[b] | 0.02327 ± 0.00663[b] |
| $D_P$ | 0.2467 ± 0.0006[b] | 2.82 ± 0.28[a] | 5.764 ± 0.360[b] | 0.02104 ± 0.00132[b] |
| *WMAP* | 0.2475 ± 0.0006[b] | 2.49 ± 0.22[b] | 6.225 ± 0.340[b] | 0.02273 ± 0.00124[a] |

[a]Observed value.
[b]Predicted value.

9. NON STANDARD PHYSICS OR NEW PHYSICS

The H II regions $Y_P$ value constrains non standard physics. Non standard physics has been discussed by many authors, a seminal paper on this subject was presented by Dirac (61), three recent papers on the $Y_P$ relevance have been presented by Cyburt et al. (62) an Coc et al. (63), and Gassner & Lesch (64), and general reviews on the variability of fundamental constants provided by astronomical data have been presented by Uzan (65) and García-Berro, Isern, & Kubyshin (66).

The $Y_P$ value derived from H II regions provides important constrains to some predictions of non-standard cosmologies such as: (a) the number of light neutrino families at the time of BBN, (b) the presence of decaying particles during BBN which could have affected the production of light elements, (c) time variations in fundamental constants like Newton's constant G, the fine structure constant α, the proton to electron mass ratio, the neutron lifetime, and the neutron proton mass difference, (d) the possible time variation of the Higgs field, and (e) the presence of vacuum energy in the tracker regime during BBN.

In what follows we will discuss the restrictions on the number of light neutrino families produced by the combination of $Y_P$ and $D_P$.

## 9. Number of light neutrino families

The best way to constrain the number of light neutrino families is provided by combining the observed $Y_P$ and $D_P$ values. Figure 7 presents $Y_P$ as a function of $D_P$ from BBN predictions for three choices of $N_v$, the number of light neutrino families. From this figure and the results for $Y_P$ and $D_P$ presented in Table 3 we obtain $N_{eff} = 2.99 \pm 0.23(68\%)$, where $N_{eff}$ is the effective number of neutrino families. For $\tau_n = 878.5$ sec (59) we obtain $N_{eff} = 3.12 \pm 0.23(68\%)$ and for $\tau_n = 881.9$ sec (60) we obtain $N_{eff} = 3.06 \pm 0.23(68\%)$.

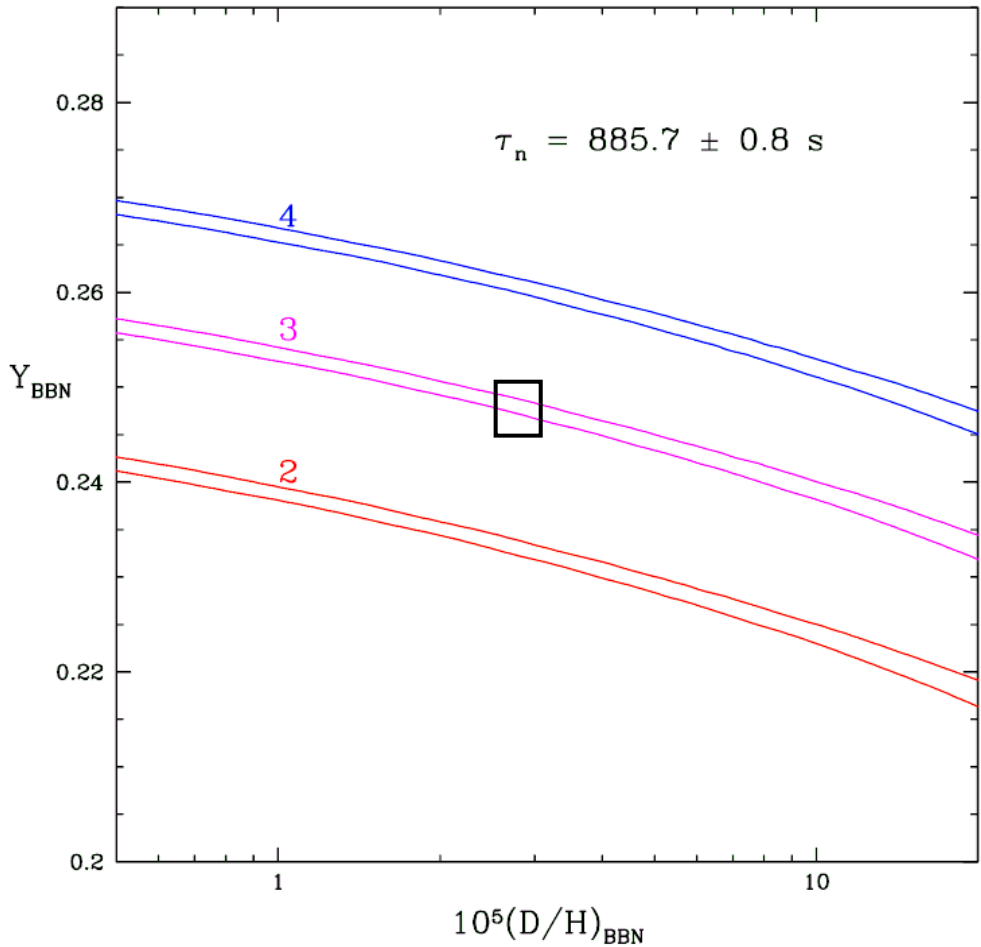

Figure 7 The BBN-predicted primordial $^4$He mass fraction, $Y_P$, as a function of the BBN-predicted deuterium abundance, $D_P$, for three choices of $N_v$. The width of the band represents the theoretical uncertainty, largely due to that of the neutron lifetime, $\tau_n$, presented in the figure. Figure from Steigman (22) under the assumption that $\tau_n = 885.7 \pm 0.8$ sec (31). The error in $\tau_n$ might be considerably larger than the presented in this figure, see text. I have added to this figure the square that comes from the $Y_P$ and $D_P$ values by Peimbert, Luridiana, & Peimbert (29) and by O'Meara et al. (58) respectively, presented in Table 3.

It is also possible from the *WMAP* observations to present an estimate of $N_{eff}$. From Figure 8 due to Komatsu et al. (67) it can be seen that *WMAP* by itself does not provide a good restriction for $N_{eff}$, but by combining the *WMAP* results with those derived from supernovae of Type Ia, baryon acoustic oscillations, and the Hubble parameter, $H_0$, derived from the Hubble Space Telescope, Komatsu et al. (67) find that $N_{eff} = 4.4 \pm 1.5(68\%)$.

Based on the production of the Z particle by electron-positron collisions in the laboratory it is found that $N_{lab} = 2.984 \pm 0.008(68\%)$ (68). According to Mangano et al. (69) this value corresponds to $N_{eff} = 3.04$ due to a partial heating of neutrinos produced by electron-positron annihilations during BBN. This $N_{eff}$ value is presented in Figure 8 as the Standard Value.

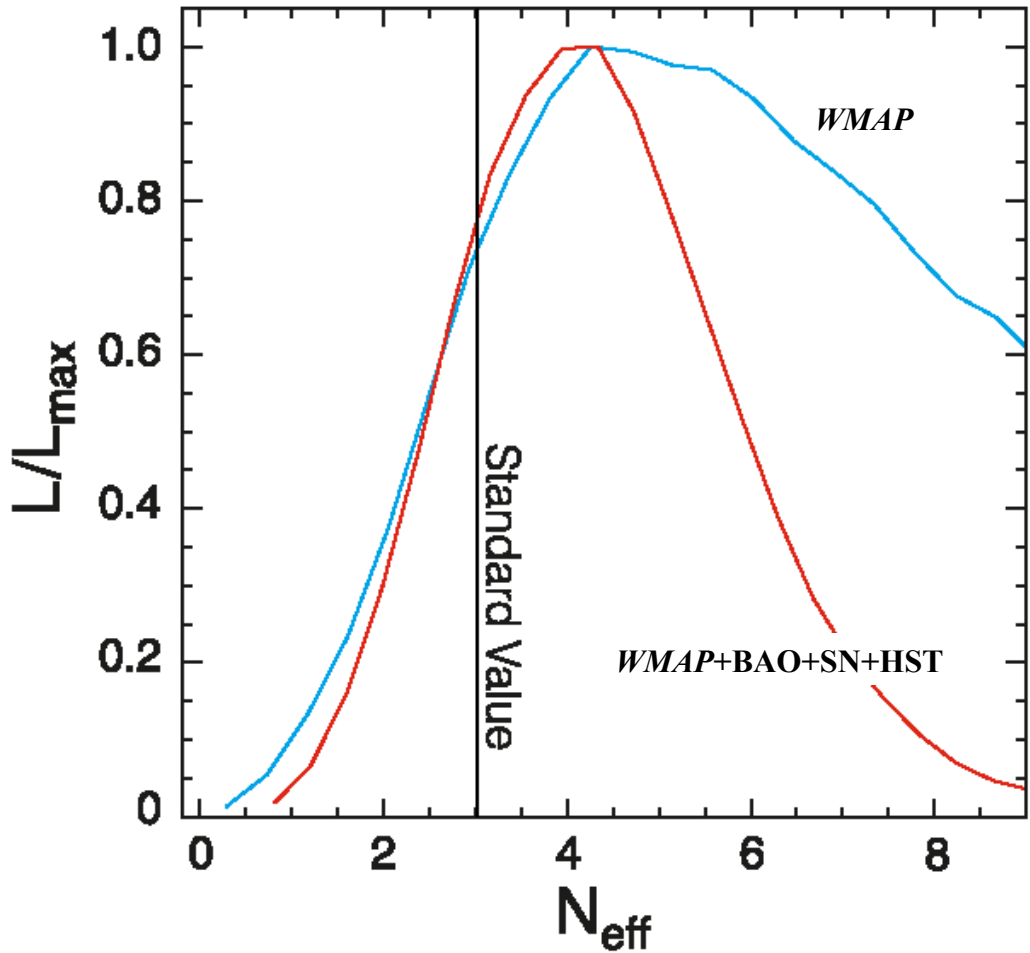

Figure 8. Constraint on the effective number of neutrino species, $N_{eff}$. One dimensional marginalized distribution of $N_{eff}$ from *WMAP*-only, *WMAP*, and the combination of the *WMAP* results with those of the Baryon Acoustic Oscillations in the distribution of galaxies, the distance measurements from Type Ia supernovae and the Hubble Space Telescope value for $H_0$, the Hubble parameter, that amounts to $72 \pm 8$

km/s/Mpc, *WMAP* + BAO + SN + HST. The standard value for $N_{eff}$ = 3.04 is shown by the vertical line. Figure from Komatsu et al. (67).

## 10. CONCLUSIONS AND OUTLOOK

During the last 50 years the determination of $Y_P$ has been very important for the study of cosmology, stellar evolution, and the chemical evolution of galaxies. To determine $Y_P$ it is necessary to determine accurate atomic parameters and the physical conditions inside ionized gaseous nebulae.

During the last five decades the accuracy of the $Y_P$ determination has increased considerably, and during the last two decades the differences among the best $Y_P$ determinations have been due to systematic effects. These systematic effects have been gradually understood, particularly during the last few years.

The best $Y_P$ determination available, that by Peimbert, Luridiana, & Peimbert (29), is in agreement with the $D_P$ determination and with the *WMAP* observations under the assumption of SBBN. The errors in the $Y_P$ determination are still large and there is room for non-standard physics or new physics.

From the $Y_P$ by Peimbert, Luridiana, & Peimbert (29), the $D_P$ by O'Meara et al (58), and BBN it is found that for $\tau_n$ = 881.9 ± 1.6 sec the number of effective neutrino families, $N_{eff}$, is equal to 3.06 ± 0.23(68%). Based on the production of the Z particle by electron-positron collisions in the laboratory and taking into account the partial heating of neutrinos produced by electron-positron annihilations during BBN Mangano et al. (69) find that Neff = 3.04. The $N_{eff}$ value derived from $Y_P$, $D_P$, and BBN is in excellent agreement with the value derived by Mangano et al (69).

The accuracy of the $N_{eff}$ determination based on $Y_P$, $D_P$, and BBN is considerably higher than that derived from *WMAP* + BAO + SN + HST.

To improve the accuracy of the $Y_P$ determination the following steps should be taken in the near future: (a) to obtain new observations of high spectral resolution of metal poor H II regions, those with 0.0005 < Z < 0.001 to reduce the effect of the collisional excitation of the Balmer lines that for the present day $Y_P$ determinations is the main source of error; (b) to determine the temperature of the H II regions based on the Balmer continuum to a Balmer line ratio, or based on a large number of He I lines observed with high accuracy, and to try to avoid the use of $T$(O III) temperatures that weigh preferentially the regions of higher temperature than the average one, effect that artificially increases the $Y_P$ determinations; (c) additional efforts should be made to understand the mechanisms that produce temperature variations in giant H II regions and once they are understood they should be incorporated into phtoionization models; (d) the He I recombination coefficients should be computed again with an accuracy higher than that of the last two determinations; (e) similarly a new determination of the neutron lifetime is needed to sort out the difference between the result obtained by Arzumanov et al. (31) and the result obtained by Serebrov et al. (59).

It is a pleasure to acknowledge that many of the ideas mentioned in this article were the product of my collaboration with: L. Carigi, C. Esteban, J. Fierro, V. Luridiana, C. R. O'Dell, A. Peimbert, M. Peña, M. T. Ruiz, and S. Torres-Peimbert. It is also a pleasure to acknowledge many fruitful discussions carried out over the years on the subject of the primordial helium abundance with: L. H. Aller, G. Ferland, D. E. Osterbrock, B. E. J. Pagel, R. L. Porter, M. J. Seaton, E. D. Skillman, and G. Steigman. This work was partly supported by the CONACyT grant 46904.

REFERENCES

1. Boesgaard AM, Steigman G. Big bang nucleosynthesis: theories and observations. Annu. Rev. Astron. & Astrophys. 1985; 23: 319-378

2. Olive KA, Steigman G, Walker TP. Primordial nucleosynthesis: theory and observations. Phys. Rep. 2000; 333: 389-407.

3. Steigman G. Primordial nucleosynthesis in the precision cosmology era. Annu. Rev. Nucl. Part. Sci. 2007; 57: 463-491.

4. Peimbert M, Torres-Peimbert S. Peebles's Analysis of the Primordial Fireball. Astrophys. J. Centennial Issue 1999; 525C: 1143-1146.

5. Burbidge GR. Cosmic helium. Comm. Astrophys. Space Phys. 1969; 1: 101-106.

6. Danziger IJ. The cosmic abundance of helium. Annu. Rev. Astron. Astrophys. 1970; 8: 161-178.

7. Osterbrock DE, Rogerson JB. The helium and heavy-element content of gaseous-nebulae and the Sun. Pub. Astron. Soc. Pacific 1961; 73: 129-134.

8. Aller LH. The abundance of the elements. Interscience Publishers, New York 1961.

9. O'Dell CR, Peimbert M, Kinman TD. The planetary nebula in M15. Astrophys. J. 1964; 140: 119-129.

10. Sargent WLE, Searle L. Spectroscopic evidence on the helium abundance of stars in the galactic halo. Astrophys. J. 1966; 145: 652-654.

11. Greenstein JL, Munch G. The weakness of helium lines in globular clusters and halo B stars. Astrophys. J. 1966; 146: 618-620.

12. Greenstein JL, Truran JW, Cameron AGW. Helium deficiency in old halo B stars. Nature 1967; 213: 871-873.

13. Hoyle F, Tayler RJ. The mystery of the cosmic helium abundance Nature. 1964; 203: 1108-1110.

14. Peebles PJE. Primordial helium abundance and the primordial fireball. Astrophys. J. 1966; 146: 542-552.

15. Wagoner RV, Fowler WA, Hoyle F. On the synthesis of elements at very high temperatures. Astrophys. J. 1967; 148:3-49.

16. Searle L, Sargent WLW. The observational status of cosmological helium. Comm. Astrophys. Space Phys. 1972; 4: 59-64.

17. Searle L, Sargent WLW. Inferences from the composition of two dwarf blue galaxies . Astrophys. J. 1972; 173: 25-33.

18. Peimbert M. Planetary nebula V: On the planetary nebula in M15. Mem. Soc. Roy. Sci. Liege, 6e serie 1973, 5: 307-317.

19. Peimbert M, Torres-Peimbert, S. Chemical composition of H II regions in the Large Magallanic Cloud and its cosmological implications. Astrophys. J. 1974; 193: 327-333.

20. Peimbert M, Torres-Peimbert S. Chemical composition of H II regions in the Small Magellanic Cloud and the pregalactic helium abundance. Astrophys. J. 1976; 203: 581-586.

21. Lequeux J, Peimbert M, Rayo JF, Serrano A, Torres-Peimbert S. Chemical composition and evolution of irregular and blue compact galaxies. Astron. & Astrophys. 1979; 80: 155-166.

22. Steigman G. Primordial alchemy: from the Big Bang to the present universe. Course of lectures at the XIII Canary Islands Winter School of Astrophysics; Cosmochemistry: the melting pot of elements.(November 19 - 30, 2001; Tenerife, Canary Islands, Spain) 2002; astro-ph 0.08186.

23. Peimbert A. The chemical composition of the 30 Doradus nebula derived from Very Large Telescope echelle spectrophotometry. Astrophys. J. 2003; 584: 735-750.

24. Pagel BEJ, Simonson EA, Terlevich RJ, Edmunds MJ. The primordial helium abundance from observations of extragalactic H II regions. MNRAS 1992; 255: 325-345.

25. Peimbert M, Peimbert A, Ruiz MT. The chemical composition of the Small Magellanic Cloud H II region NGC 346 and the primordial helium abundance. Astrophys. J. 2000; 541: 688-700.

26. Luridiana V, Peimbert A, Peimbert M, Cerviño M. The effect of collisional enhancement of Balmer lines on the determination of the primordial helium abundance. Astrophys. J. 2003; 592: 846-865.

27. Izotov YI, Thuan TX. Systematic effects and a new determination of the primordial abundance of $^4$He and d$Y$/d$Z$ from observations of blue compact galaxies. Astrophys. J. 2004; 602: 200-230

28. Izotov YI, Thuan TX, Stasińska G. The primordial abundance of $^4$He: a self consistent empirical analysis of systematic effects in a large sample of low metallicity H II regions. Astrophys. J. 2007; 662: 15-38.

29. Peimbert M, Luridiana V, Peimbert A. Revised primordial helium abundance based on new atomic data. Astrophys. J. 2007; 666: 636-646.

30. Dunkley J, et al. Five-year Wilkinson Microwave Anisotropy Probe (*WMAP*) observations: likelihoods and parameters from the *WMAP* data. Astrophys. J. 2008; submitted: arXiv0803.0586.

31. Arzumanov S, et al. Neutron life time value measured by storing ultracold neutrons with detection of inelastically scattered neutrons. Physics Letters B 2000 ; 483:15-22.

32. González Delgado RM, Leitherer C, Heckman TM. Synthetic spectra of H Balmer and He I absorption lines. II. Evolutionary synthesis models for starburst and poststarburst galaxies Astrophys. J. Supp. 1999; 125: 489-509.

33. González Delgado RM, Cerviño M, Martins LP, Leitherer C, Hauschildt PH. Evolutionary stellar population synthesis at high spectral resolution: optical wavelengths. MNRAS 2005; 357: 945-960.

34. Porter RL, Bauman RP, Ferland GJ, MacAdam KB. Theoretical He I emissivities in the Case B approximation. Astrophys. J. 2005; 622: L73-L75.

35. Porter RL, Ferland GJ, MacAdam KB. He I emission in the Orion Nebula and implications for primordial helium abundance. Astrophys. J. 2007; 657: 327-337.

36. Smits DP. Theoretical He I line intensities in low-density plasmas. MNRAS 1996; 278: 683-687.

37. Benjamin RA, Skillman ED, Smits DP. Improving predictions for helium emission lines. Astrophys. J. 1999; 514: 307-324.

38. Anderson H, Ballance CP, Badnell NR, Summers HP. An R-matrix with pseudostates approach to the electron-impact excitation of H I for diagnostic applications in fusion plasmas. J. Phys. B. 2000; 33: 1255-1262.

39. Anderson H, Ballance CP, Badnell NR, Summers HP. CORRIGENDUM: an R-matrix with pseudo-states approach to the electron-impact excitation of H I for diagnostic applications in fusion plasmas. J. Phys. B. 2002; 35: 1613-1615.

40. Olive KA, Skillman ED. A realistic determination of the error on the primordial helium abundance: steps toward nonparametric nebular helium abundances. Astrophys. J. 2004; 617: 29-49.

41. Peimbert A, Peimbert M, Luridiana V. Temperature bias and the primordial helium abundance determination. Astrophys. J. 2002; 565: 668-680.

42. Peimbert M, Luridania V, Peimbert A, Carigi L. On the primordial helium abundance and the $\Delta Y/\Delta O$ ratio. In From Stars to Galaxies: Building the Pieces to Build

Up the Universe. A Vallenari, R Tantalo, L Portinari, A. Moretti, eds., ASP Conference Series Vol. 374, 2007; pp. 81-88.

43. Relaño M, Peimbert M, Beckman J. Photoionization models of NGC 346. Astrophys. J. 2002; 564: 704-711

44. Bauman RP, Porter RL, Ferland GJ, MacAdam KB. J-resolved He I emission predictions in the low-density limit. Astrophys. J. 2005; 628: 541-554.

45. Peimbert M. Temperature determinations of H II regions. Astrophys. J. 1967; 150: 825-834.

46. Stasinska G. A grid of model H II regions for extragalactic studies. Astron. Astrophys. Suppl. 1990; 83: 501-538.

47. Moore BD, Hester JJ, Dufour RJ. Systematic errors in elemental abundances derived from nebular spectra. Astron. J. 2004; 127: 3484-3492.

48. Carigi L, Colín P, Peimbert M, Sarmiento A. Chemical evolution of irregular and blue compact galaxies. Astrophys. J. 1995; 445: 98-107.

49. Carigi L, Colín P, Peimbert M. Dark matter and the chemical evolution of irregular galaxies. Astrophys. J. 1999; 514: 787-797.

50. Carigi L, Colín P, Peimbert M. Chemical and photometric evolution of the Local Group galaxy NGC 6822 in a cosmological context. Astrophys. J. 2006; 644: 924-939

51. Kroupa P, Tout CA, Gilmore G. The distribution of low-mass stars in the Galactic disc. MNRAS 1993; 262: 545-587.

52. Carigi L, Peimbert M. The helium and heavy elements enrichment of the interstellar medium predicted by models for our Galaxy. In preparation 2008.

53. Carigi L, Peimbert M, Esteban C, García-Rojas J. Carbon, nitrogen, and oxygen Galactic gradients: a solution to the carbon enrichment problem. Astrophys. J. 2005; 623: 213-224.

54. Maeder A. Stellar yields as a function of initial metallicity and mass limit for black hole formation. Astron. Astrophys. 1992; 264: 105-120.

55. Hirschi R, Meynet G, Maeder A. Yields of rotating stars at solar metallicity. Astron. Astrophys. 2005; 433: 1013-1022.

56. Steigman G. The cosmological evolution of the average mass per baryon. J. Cosmol. Astropart. Phys. 2006; 10: 16-22.

57. Steigman G. Primordial nucleosynthesis: successes and challenges. Int. J. Mod. Phys E 2006; 15: 1-35.

58. O'Meara JM, Burles S, Prochaska JX, Prochter GE. The deuterium-to-hydrogen abundance ratio toward the QSO SDSS J155810.16-003120.0 Astrophys. J. 2006; 649: L61-L65.

59. Serebrov A, et al. Measurement of the neutron lifetime using a gravitational trap and a low-temperature Fomblin coating. Physics Letters B. 2005; 605: 72-78.

60. Mathews GJ, Kajino T, Shima T. Big bang nucleosynthesis with a new neutron lifetime. Physical Review D 2005; 71: 021302.

61. Dirac PAM. The cosmological constants. Nature 1937; 139: 323.

62. Cyburt RH, Fields BD, Olive KA, Skillman ED. New BBN limits on physics beyond the standard model from $^4$He. Astropart. Phys.2005; 23: 313-323.

63. Coc A, Nunes NJ, Olive KA, Uzan J-P, Vangioni E. Coupled variations of fundamental couplings and primordial nucleosynthesis. Phys. Rev. D 2007; 76: 023511.

64. Gassner JM, Lesch H. Primordial $^4$He abundance constrains the possible time variation of the Higgs vacuum expectation value. Int. J. Theor. Phys. 2008; 47: 438-445.

65. Uzan J-P. The fundamental constants and their variation:observational status and theoretical motivations. Rev. Mod. Phys. 2003; 75: 403-455.

66. García-Berro E, Isern J, Kubyshin YA. Astronomical measurements and constraints on the variability of fundamental constants. Astron. Astrophys. Rev. 2007; 14: 113-170.

67. Komatsu E, et al. Five-year Wilkinson Microwave Anisotropy Probe (*WMAP*) observations: cosmological interpretation. Astrophys. J. 2008; submitted: ArXiv0803.0647.

68. Yao W-M, et al. Review of Particle Physics. J. Phys. G. Nucl. Part. Phys. 2006; 33: 1- 1232.

69. Mangano G, Miele G, Pastor S, Peloso M. A precision calculation of the effective number of cosmological neutrinos. Phys. Let. B 2002; 534: 8-16.